# An electrically tunable terahertz plasmonic device based on shape memory alloys and liquid metals

*Hui Zhou,[1, 2*#] Ting Zhang,[1 2*#] and Ajay Nahata[2]*

[1]Department of Physics and Astronomy, University of Utah, Salt Lake City, UT, 84112, USA
[2]Department of Electrical and Computer Engineering, University of Utah, Salt Lake City, UT, 84112, USA
# These authors contributed equally
*E-mail: hui.zhou@utah.edu, ting.zhang@utah.edu

**Abstract:** An electrically tunable terahertz (THz) plasmonic device is designed and fabricated using liquid metals (eutectic gallium indium – EGaIn) and shape memory alloy wires (Flexinol). The liquid metal is injected into the voids of a poly(dimethyl) siloxane (PDMS) microfluidic mold forming a periodic array of subwavelength apertures, while the wires are inserted into the elastomer below the metal plane. When a DC voltage is applied to the wires, they contract via Joule heating, reducing the aperture periodicity and blue-shifting the transmission resonances of the device. When the voltage is removed, the wires cool and elongate back to their original length, allowing the transmission spectrum to return to its original state. The magnitude of this change depends upon the applied voltage. The device is shown to thermally cycle between the relaxed state and the fully contracted state reproducibly over at least 500 thermal cycles. The asymmetric geometry of the device and the contraction process yield transmission properties that are unexpected: two closely spaced resonances, where both resonances correspond to the same scattering indices, and an increase in the transmission amplitude of the lowest order resonance upon contraction. We use numerical simulations to understand these features.

## 1. Introduction

Artificially structured materials, which encompass architectures that are commonly categorized as either metamaterials [1] or plasmonic structures [2], have elicited great interest over the last two decades, because of their promise in allowing for the optical response of the medium to be engineered. The latter topic involves media that support surface plasmon-polaritons (SPPs) at metal-dielectric interfaces and has enabled a broad range of applications including miniaturized photonic circuits [3], super-resolution imaging [4] and filters [5,6]. In the THz spectral range, plasmonic structures are particularly interesting because conventional metals exhibit extremely low loss over the length scales relevant to most devices [7]. This results from the high conductivity of most metals, but also limits their applicability to passive devices, since their conductivities cannot be easily altered. While such devices are extremely useful, it is also highly desirable to create structures that are active, such that the THz response can be modulated or continuously tuned.

In the long wavelength regime, which includes the THz spectral band that ranges from 100 GHz to 10 THz [8], a broad range of exotic metals have been shown to support relatively low loss propagation of SPPs. Such materials are attractive because they allow for dynamic changes in the optical response. These include semiconductors [9,10], graphene [11,12], superconductors [13,14] and phase-change materials such as $VO_2$ [15], which can have their dielectric properties altered via optical, electrical or thermal excitation. Analogous capabilities have also been shown with metamaterials [16,17]. Such changes generally modify the conductivity of the material and, thus, only modulate the amplitude of the response function.

Tuning of the spectral response, on the other hand, requires either a change in the physical dimensions of the structure or a change in the refractive index of the surrounding dielectric medium. Both approaches have been used with metamaterial devices, where changes have been made to the unit cell properties [18-22]. Even then, recent examples of frequency tuning have been limited to approximately 5% [22]. In the case of plasmonic structures, where the response is associated with the propagation properties of surface plasmon-polaritons (SPPs) over larger length scales, there are few demonstrations of frequency tuning [23].

In this submission, we describe a THz plasmonic device in which the transmission resonances can be continuously tuned electrically by changing the length of the metallic structure. The device is based on a periodic array of subwavelength apertures that combines the advantageous attributes of both liquid metals and shape memory alloys. The former material can be stretched and otherwise deformed without buckling or creating cracks or voids that penetrate the entire metal layer, while the latter material can change shape and be returned to its original form by simply changing the temperature. The specific device we have created is defined by injecting a liquid metal (eutectic gallium indium – EGaIn [24]) into a poly(dimethyl) siloxane (PDMS) microfluidic mold. We have previously shown that EGaIn is attractive for both THz plasmonic and metamaterial applications [25-27], despite the fact that its DC conductivity is smaller than that of stainless steel and lead [28]. We then insert shape memory alloy (Flexinol) wires into the device [29]. These wires have been trained to contract via Joule heating and relax back to their original length when cooled. We use THz time-domain spectroscopy to measure the transmission properties of the device as a function of the applied voltage and numerical simulations to validate and further understand the results.

## 2. Results and Discussion

In **Figure 1a**, we show a schematic diagram of the device. The PDMS microfluidic mold is composed of a lower section that is 1750 μm thick and an upper section that is 315 μm thick. The rationale for this asymmetric PDMS geometry is discussed below. The periodic aperture array consists of a 30 μm thick region of EGaIn that is injected into the microfluidic mold between the two PDMS layers, where the apertures are formed by PDMS cylinders that are 110 μm in diameter and 30 μm tall (i.e. they connect the two PDMS layers and EGaIn fills in the region surrounding the pillars) with a periodic spacing of 215 μm in a 40 x 40 square lattice. The lower PDMS region contains two 150 μm diameter Flexinol wires, spaced by ~3.9 mm, that lie parallel to an axis of the aperture array. Flexinol is composed of a nickel-titanium alloy [30,31] that has been trained to contract upon heating and elongate back to its original length when subject to a bias force upon cooling. These wires are inserted into the PDMS mold such that they do not come in contact with the liquid metal layer and do not adversely affect the operation of the device (see Supplementary Information).

The 150 μm diameter Flexinol wires exhibit a resistance of 50 Ω/m at room temperature, corresponding to a resistance of 10 Ω for the two 10 cm long wires connected in series. Thus, the application of only a few volts DC causes sufficient heating to allow the wires to contract. When the applied voltage is removed, the wires quickly cool and return to their original length. Before insertion into the PDMS, we ensured that the wires met the manufacturer specifications in terms of dimensional changes [29]. The asymmetric PDMS geometry was used to balance considerations of device planarity and THz absorption. Specifically, a thickness of 1.75 mm for the lower PDMS section was used to ensure that the device remained planar as the Flexinol wires contracted (see

Supplementary Information for images). It is important to note that when the lower PDMS section was made thinner, the device would often buckle in a manner that was not uniform across the sample, making it difficult to obtain consistent results. Since the lower section was made intentionally thick to ensure that the device remained planar, the upper PDMS section was made thin to minimize the overall THz absorption; PDMS exhibits broadband attenuation of THz radiation, corresponding to reduction in the transmitted THz electric field of ~26% / mm thickness in the 0.1 – 0.5 THz range [25].

In Figure 1b and 1 c, we show images of the overall and a magnified portion of the final device. The liquid metal was injected into the sides of the device. With 4V (~1.5 W) voltage applied, the dissipated power (determined by measuring both the voltage and the current) would heat up the Flexinol wires which would contract the sample and reduce the periodicity along the wire axis. The dimension remained unchanged along the orthogonal axis. In all of the experimental results described below, the incident and measured THz radiation was polarized parallel to the wires. Thus, the arrays parameters along the orthogonal axis, which appear to be unaffected by the application of an external voltage, do not play any role in the observed response.

We measured the transmission properties of the device as a function of the voltage applied to the Flexinol wires. In **Figure 2a**, we show the terahertz transmission spectra for two values of the applied voltage: 0 V (0 W) and 4 V (~1.5 W). Both spectra consist of two closely spaced resonances, with a frequency shift upon application of the voltage. We begin by explaining the locations of the resonances. To do so, we consider the momentum matching condition rewritten to yield the resonance frequencies in terms of the array parameters [32]:

$$v_{AR} = \frac{c\sqrt{i^2+j^2}}{Pn_{SPP}}, \qquad (1)$$

where

$$n_{SPP} = \left(\frac{\varepsilon_m \varepsilon_d}{\varepsilon_m + \varepsilon_m}\right)^{1/2} \qquad (2)$$

In these equations, c is the speed of light in vacuum, i and j are integers that index the scattering resonance order [2,32], P is the aperture periodicity, $n_{SPP}$ is the effective refractive index for the propagating SPP, and $\varepsilon_m$ and $\varepsilon_d$ are the complex dielectric constants of the metal and adjacent dielectric medium. We have previously shown that the frequencies resulting from this equation correspond to the resonance dips and not to the resonance peaks [33]

Since the real and imaginary components of the complex dielectric constant of EGaIn are much larger than those of PDMS, $n_{SPP} \cong n_{THz(PDMS)}$ [9]. The measured real part of the THz refractive index for PDMS was found to be ~1.57 near 1 THz, which is consistent with earlier measurements obtained in the 0.1 – 0.5 THz frequency range [25]. Thus, the frequency associated with the lowest order (i = 0, j = ±1) resonance dip is expected to occur at 0.89 THz according to Eq. 1 for 0 V. This closely matches the observed value of 0.90 THz. However, no values of i and j yield the second resonance dip frequency of 1.03 THz. When 4 V (~1.5 W) was applied to the Flexinol, the lowest order (i = 0, j = ±1) resonance dip occurs at 0.94 THz, corresponding to 5% periodicity change, which is a reduction in the periodicity of the aperture array from 215 to 204 um. However, the properties of the second resonance, with a dip at 1.07 THz cannot be obtained using Eq. 1 once again. In addition, the maximum amplitude of the lowest order resonance unexpectedly increases slightly as the resonances shift to the right upon application of a voltage.

To understand the origin of the second resonance for both applied voltages and the change in amplitudes, we simulated the electromagnetic response of the aperture array using numerical finite-difference time-domain methods (Lumerical), using the measured device parameters noted above. As in the experiment, the incident and detected THz radiation was polarized parallel to the Flexinol wires. In Figure 2b, we show the simulated transmission spectra for both applied voltages, which is in good agreement with the experimental results. What the numerical simulations demonstrate is that the lower frequency resonance is associated with metal – lower PDMS layer. Because that PDMS layer is very thick, SPPs along the lower metal – dielectric interface see an effective $n_{SPP} \cong 1.57$. On the other hand, the effective $n_{SPP}$ for the metal – upper PDMS layer is $\cong$ 1.36. This occurs because the upper PDMS later is only 315 μm thick. We have previously shown that the 1/e out-of-plane decay length for SPPs typically corresponds to ~3-5 wavelengths in free space [7]. Thus, the evanescent properties of SPPs extend through the upper PDMS layer and into the air, necessitating an effective index that take both dielectric media into account. The small change in the aperture shape also accounts for the change in the resonance amplitudes.

Shape memory alloys exhibit hysteresis as they undergo a reversible thermally induced phase transition. In the case of Nitinol foils, we measured the hysteresis properties of the alloy by directly measuring the metal temperature as it thermally cycled [34]. In the case of the Flexinol wires, we were not able to measure the wire temperatures directly, since a portion of these wires were embedded within PDMS. However, the power dissipated in the wires is directly proportional to the temperature. In **Figure 3a**, we show the hysteresis properties by measuring the device contraction as a function of dissipation power, and the y axis is the corresponding frequency of the frequency dip which is calculated based on the resonant frequency changes for the lowest order caused by the periodicity change. The device length was measured by taking high resolution images of the structure as the voltage was varied and measuring the dimensions. The device contraction is consistent with the expected wire contraction, based on the manufacturer's data [29] and our own measurements.

In principle, bare Flexinol wires should be able to thermally cycle more than $10^6$ times if the length contraction is kept within a ~5% range, as is done here. Thus, failure of the device is unlikely to arise from the wire itself, but rather because of how the wires are attached. For example, if the wires are glued, small length changes in the wire cause it to detach from the rigid adhesive, leading to failure. Since PDMS is flexible, we do not expect such issues. In Figure 3b, we show the thermal cycling over 500 thermal cycles, based on measurement of the frequency of the dip for the lowest order resonance. It is apparent that there is no significant variation in the THz response over the entire cycling range.

Finally, we comment on alternate geometries that may allow for larger dimensional changes, corresponding to a wider tuning range. In contrast to straight wires, where the fractional contraction are limited, Flexinol wires in the form of coils have been shown to exhibit length changes that can exceed 100% along the coil axis [29]. Unfortunately, such coils cannot easily be integrated into the existing device because of how they would unwind within the elastomer. However, shape memory alloys that assume a planar zig-zag geometry should yield somewhat similar results. However, the alloy would need to undergo two-way training to enable such a capability [35]. More generally, a variety of different wire-based embodiments have been designed that allow for relatively large fractional changes. Some of these implementations may be relevant for tunable THz devices.

## 3. Conclusion

In conclusion, we have demonstrated a tunable THz plasmonic device that utilizes the advantageous properties of liquid metals and shape memory alloys. Flexinol wires inserted into the elastomeric medium of the device allow for electrical control over the device dimensions, while the EGaIn ensures that the metal contracts and relaxes without cracking or creating air gaps in the plasmonic layer. By varying the voltage across the wires, we controlled the extent to which the Flexinol wires contracted, thereby tuning the aperture periodicity and transmission spectrum. In the present geometry, we observed a ~5% tuning range, which is similar to what has been observed using other approaches [22]. Compared to electrostatic approaches, device fabrication using Nitinol wires and liquid metals is significantly simpler. Furthermore, we expect that alternate wire geometries combined with the use of elastomeric materials that allow for greater stretching ratios will allow for larger dimensional changes corresponding to larger tuning ranges.

## 4. Experimental Section

*Fabrication of device*: A silicon wafer coated with 30 µm of SU-8 2025 photoresist was spun cast, baked, and exposed to UV light using an appropriately patterned mask that served as the template for the periodic aperture array in the following soft lithography process. A PDMS pre-polymer was mixed with a curing agent using a weight ratio of 7:1, degassed, poured onto the SU-8 template, and cured for 2 hours at 60 °C. After curing, the inverse PDMS replica were peeled off and sealed with a 315 µm thick planar section of PDMS using a high voltage corona, yielding the desired microfluidic mold. EGaIn was then injected into the microchannels using a syringe to form the desired geometry. Flexinol wires were then laid on the back surface of the device and a thicker layer of PDMS was poured, yielding a lower PDMS layer that was 1.75 mm thick. This layer was cured at 55 °C for 4 hours to minimize changes in the Flexinol.

*Characterization of transmission properties*: We used an amplified ultrafast Ti: Sapphire laser as the optical source for all of the THz time-domain spectroscopy measurements. The output of the laser was split 80:20 to yield the optical pump and probe beams, respectively. Broadband THz radiation was generated using a 1 mm thick <110> ZnTe crystal. An off-axis paraboloidal mirror was used to collect and collimate the THz radiation as it propagated from the emitter to the sample, resulting in a beam that was normally incident onto the sample. A second off-axis paraboloidal mirror was used to refocus the transmitted THz radiation onto a 1 mm thick <110> ZnTe detection crystal, which allowed for coherent detection of the radiation via electro-optic sampling [36].

**Acknowledgements**

This work was supported by the NSF MRSEC program at the University of Utah under grant # DMR 1121252. This work made use of University of Utah USTAR shared facilities. We thank Brian Baker, Hao-Chieh Hsieh and Kevin Petersen for assistance with fabrication.

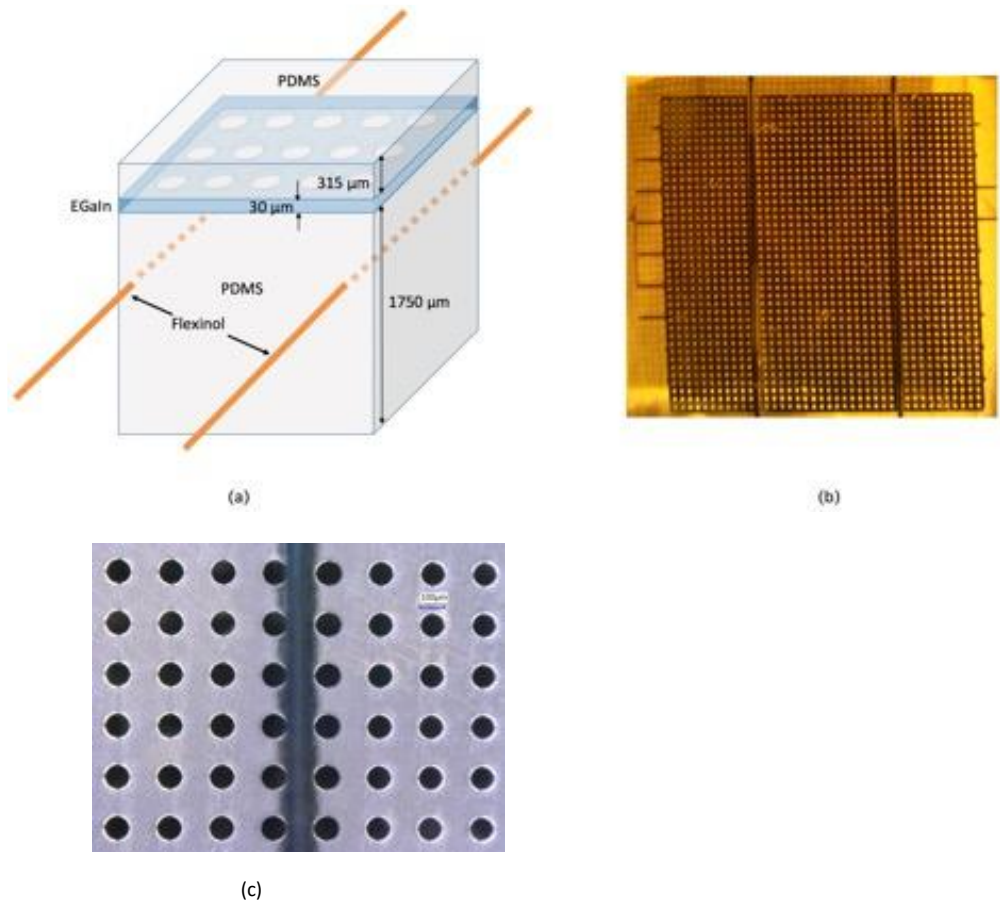

**Figure 1.** Device properties (a) Schematic diagram of the device. The device consists of a 40 x 40 array of periodically spaced circular apertures. The aperture parameters are discussed below. (b) Image of the overall device. The horizontal lines along the left and right edges of the device are excess EGaIn in the injection lines. (c) Image of a portion of the aperture array with 0 V (0 W) applied that demonstrates that the periodicity is 215 μm along both the vertical and horizontal axes. The aperture diameter is 110 μm along both the axes.

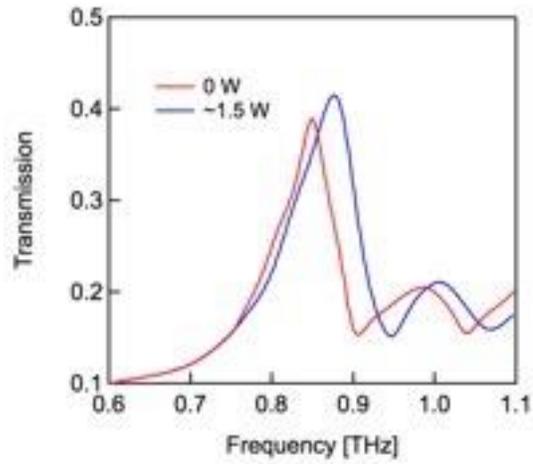

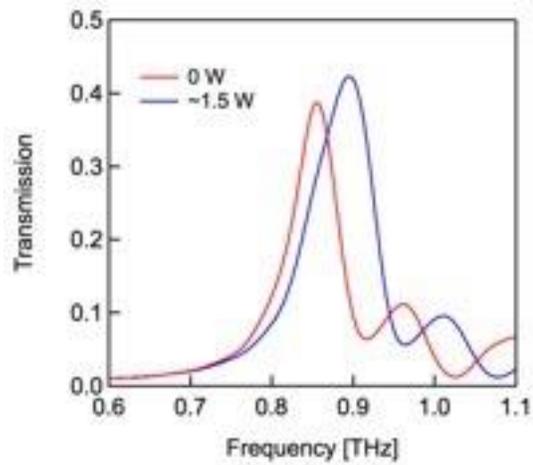

**Figure 2.** THz transmission properties for the device (a) Experimentally measured THz transmission spectrum for the device using applied voltages of 0 V (0 W) and 4 V (~1.5 W), respectively. (b) Numerically simulated THz transmission spectrum for the device using measured geometrical device parameters. The incident and measured THz radiation was polarized parallel to the Flexinol wires.

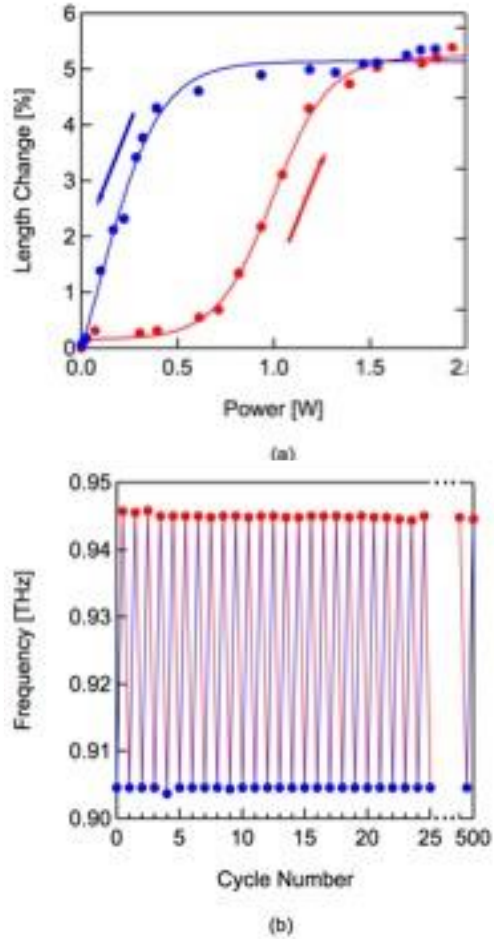

**Figure 3.** Thermal tuning of the device (a) Experimentally measured thermally-induced hysteresis of the phase transition properties of the Flexinol wires embedded in the device, measured as a function of the dissipated power (red and blue filled circles). The left axis shows the measured fractional length change in the device, while the right axis shows the corresponding measured frequency for the dip associated with the lowest order resonance with 0V and 4V applied voltages and calculated corresponding frequencies with applied voltages between 0V and 4V based on the length change. The data for increasing and decreasing voltage were independently fit to sigmoid curves (solid red and blue traces) (b) Frequency corresponding to the dip for the lowest order resonance as a function of thermal cycling between 0 V (0 W) and 4 V (~1.5 W) over 500 cycles.